\documentclass[twocolumn,showpacs,preprintnumbers,amsmath,prl,amssymb,superscriptaddress]{revtex4-1}

\pdfoutput=1

\usepackage[export]{adjustbox}
\usepackage{amsfonts}
\usepackage{amsmath}
\usepackage[makeroom]{cancel}
\usepackage[english]{babel}
\usepackage[T1]{fontenc}
\usepackage{times}
\usepackage{mathrsfs}
\usepackage{graphicx}
\usepackage{dcolumn}
\usepackage{bm}
\usepackage{wasysym}
\usepackage[colorlinks,bookmarks=true,citecolor=blue,linkcolor=red,urlcolor=blue]{hyperref}
\usepackage[tight, FIGTOPCAP, hang, raggedright, nooneline]{subfigure}
\usepackage{hyperref}
\usepackage{xcolor}
\usepackage{epsfig}
\usepackage{amssymb}
\usepackage{lipsum}

\newcommand{\subfigimg}[3][,]{%
  \setbox1=\hbox{\includegraphics[#1]{#3}}
  \leavevmode\rlap{\usebox1}
  \rlap{\hspace*{0pt}\raisebox{\dimexpr\ht1+0\baselineskip}{#2}}
  \phantom{\usebox1}
  }
  
 \definecolor{Green}{RGB}{80,182,0}

\usepackage{color}

\newcommand{\la}{\langle}
\newcommand{\ra}{\rangle}

\graphicspath{./Images/}
\usepackage{epstopdf}

\begin{document}
\title{Disorder-Free Localization}
\author{A.~Smith}
\email{as2457@cam.ac.uk}
\affiliation{T.C.M. group, Cavendish Laboratory, J.~J.~Thomson Avenue, Cambridge, CB3 0HE, United Kingdom}
\author{J.~Knolle}
\affiliation{T.C.M. group, Cavendish Laboratory, J.~J.~Thomson Avenue, Cambridge, CB3 0HE, United Kingdom}
\author{D.~L.~Kovrizhin}
\affiliation{Rudolf Peierls Centre for Theoretical Physics, 1 Keble Road, Oxford, OX1 3NP, United Kingdom}
\affiliation{NRC Kurchatov institute, 1 Kurchatov Square, 123182, Moscow, Russia}
\author{R.~Moessner}
\affiliation{Max Planck Institute for the Physics of Complex Systems, N\"{o}thnitzer Stra{\ss}e 38, 01187 Dresden, Germany}
\date{\today}

\begin{abstract}
The venerable phenomena of Anderson localization, along with the much more recent many-body localization, both depend crucially on the presence of disorder. The latter enters either in the form of quenched disorder in the parameters of the Hamiltonian, or through a special choice of a disordered initial state. Here we present a model with localization arising in a very simple, completely  translationally invariant quantum model, with only local interactions between spins and fermions. By identifying an extensive set of conserved quantities, we show that the system generates purely dynamically its own disorder, which gives rise to localization of fermionic degrees of freedom. Our work gives an answer to a decades old question whether quenched disorder is a necessary condition for localization. It also offers new insights into the physics of many-body localization, lattice gauge theories, and quantum disentangled liquids. 
\end{abstract}

\maketitle

The study of localization phenomena -- pioneered in Anderson's seminal
work on the absence of diffusion in certain random
lattices~\cite{Anderson1958} -- is receiving redoubled attention in
the context of the physics of interacting systems showing many-body
localization~\cite{Basko2006,Vasseur2016,Nandkishore2015}.
While in these systems the presence of quenched disorder plays a central
role, suggestions for interaction-induced localization in disorder-free systems appeared
early in the context of solid Helium~\cite{Kagan}. However, all of these
are limited to settings having inhomogeneous initial
states~\cite{Yao2014,Schiulaz2015}. Whether quenched disorder is a general 
precondition for
localization has  remained an open question. Here, we provide 
an explicit example to demonstrate that a disorder-free
system can generate its own randomness dynamically,
which leads to localization in one of its subsystems. Our model is exactly soluble,
thanks to an extensive number of conserved quantities, which we identify, 
allowing access to
the physics of the long-time limit. The model can be extended while preserving
its solubility, in particular towards investigations of
disorder-free localization in higher dimensions.

Localization phenomena are often diagnosed, in experiment and simulation, via the dynamical response to a global quantum quench. The underlying idea is to examine if a system thermalizes, thereby losing memory of the initial state, or whether this memory persists in the long-time limit~\cite{Enss2016,Schreiber2015,Yao2014,Schiulaz2015}. Some of the simple initial states used in these diagnostics exhibit density modulations, e.g., in
the form of a periodic density-wave pattern, or a density imbalance, with two halves of the system  separated by a `domain wall'.  The latter setup was exploited in the experimental identification of the many-body localization (MBL) transition~\cite{Choi2016}. In this experiment a complete domain-wall melting was observed in the ergodic phase, while the density imbalance remained in the localized phase at long times, showing exponential tails set by the localization length~\cite{Hauschild2016}. Another useful localization diagnostic, which does not require  inhomogeneous initial states, is based on examining deviations from linearity in the light-cone spreading of correlations after a quantum quench~\cite{Essler2016,Znidaric2008}.

In translationally invariant systems, initial state inhomogeneity has been a precondition for the emergence of 
localization. For instance, in models containing a mixture of interacting heavy and light particles~\cite{Schiulaz2015,Yao2014}, the heavy particles play a role of quasi-static disorder for the light particles. However, these models show only \textit{transient} sub-diffusive behaviour, which ultimately gives way to ergodicity at long times~\cite{Yao2014,Papic2015}. They were dubbed quasi-MBL. Related attempts concern localization in translationally invariant quantum versions of classical glassy models~\cite{Horssen2015}, whose behaviour, so far, cannot be differentiated from quasi-MBL, due to limitations on system sizes available in numerical simulations. 
Our work may also provide a new perspective on quantum disentangled
liquids (QDL)~\cite{Grover2014,Veness2016,Garrison2016} which are characterized by the lack of equilibration in one or more of the components of the liquid.
Beyond this, only the construction of single-particle hopping problems with entirely flat dispersions, so that the group velocities of any wave-packet vanishes, has succeeded in stopping particles from 
moving~\cite{PhysRevB.34.3625,PhysRevLett.81.5888}.

The model which we present here exhibits localization of \textit{purely dynamical origin}. This localization is induced by a quantum quench, and we use the abovementioned standard diagnostics to examine its nature. We stress that our model is entirely {\it disorder-free}, namely both the Hamiltonian, and -- crucially -- initial states do not require any quenched disorder. On the technical side we can analyse the nature of dynamically generated randomness, thanks to an extensive set of conserved quantities. We show that time evolution after a quantum quench  is described by a dual Hamiltonian with binary disorder. This allows us to develop an efficient numerical algorithm to access  system sizes far beyond the localization length, allowing us to conclude that the phenomenology detailed below is representative of the thermodynamic limit.

The paper is structured in the following way. First, we introduce the model and define the quench protocols we consider. Second, we identify the conserved charges and how they are associated with the dynamically-generated randomness. Third, we present our results for the fermionic and the spin subsystems of our model. For the fermionic subsystem we can extract a length scale which we compare with the relevant single-particle localization length. Finally, we close with a discussion where we make connections to related models and progress currently being made in the field. An outline of the numerical methods we use is referred to the Suplementary Material.

\begin{figure}[t]
	\centering
	\includegraphics[width=0.9\columnwidth]{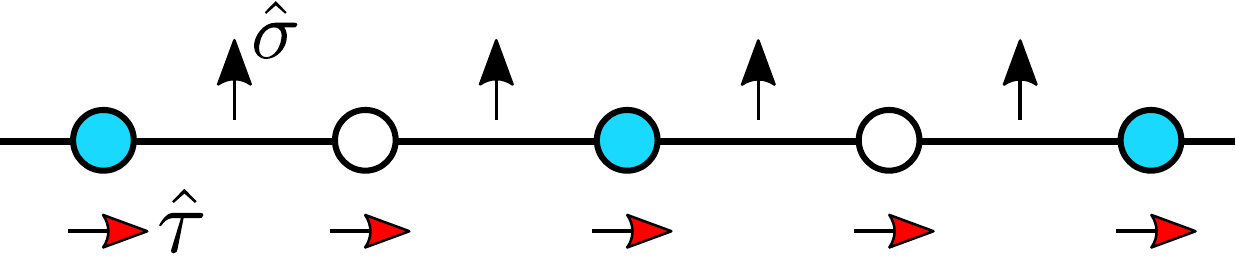}
	\caption{Schematic picture of the model. The signs of nearest-neighbour hopping for spinless fermions (blue circles) are determined by the $z$-components of $\sigma$-spins (black arrows) living on the bonds. Dual $\tau$-spins (red arrows, see Supplementary Material) are shown in the configuration corresponding to the $\sigma$-spins.}\label{fig: model}
\end{figure}

\textit{Model.} We study a 1D lattice model of spinless fermions, $\hat{f}_i$, which are coupled via spins-1/2, $\hat{\sigma}_{i,i+1}$, positioned on the bonds. The model is described by the Hamiltonian (Fig.~\ref{fig: model}),
\begin{equation}\label{eq: Hamiltonian}
\hat{H} = -J\sum_{\langle ij\rangle} \hat{\sigma}^z_{i,j} \hat{f}^\dag_{i} \hat{f}_{j}
- h \sum_{i} \hat{\sigma}^x_{i-1, i} \hat{\sigma}^x_{i,i+1} .
\end{equation}
Here $J$ and $h$ denote  fermion tunnelling strength and  Ising coupling, respectively. 
In the following we discuss dynamics induced by the Hamiltonian (\ref{eq: Hamiltonian}) on initial states with all bond spins aligned with the $z$-axis, and the fermions prepared in a Slater determinant. We consider three distinct examples of the latter: (i) domain wall $|1\ldots111000\ldots0\ra$; (ii) density wave $|\ldots10101\ldots\ra$, and (iii) {\it translationally invariant} ground state of the Hamiltonian (\ref{eq: Hamiltonian}) at $h=0$. 

\emph{Dynamically-generated randomness}. The model posses\-ses an extensive set of conserved quantities
 $\left\{q_j\right\}$ identified through the duality mapping,
  known from the Ising model~\cite{Kramers1941,Fradkin1978}. This holds for {\it arbitrary} initial fermion states. In the subspace fixed by a particular set of $\{{q}_j\}=\pm1$ the Hamiltonian (\ref{eq: Hamiltonian}) assumes a simple non-interacting form 
\begin{equation}\label{eq: hopping H}
\hat{H}_{\{q_j\}} = -J\sum_{\langle ij\rangle} \hat{c}^\dag_i \hat{c}_{j}  + 2h\sum_{j} q_j (\hat{c}^\dag_j \hat{c}_j-1/2),
\end{equation}
a tight-binding model with a binary potential given by the charge sector $\{q_j\}$. Note that, despite the simplicity
of equation~\eqref{eq: hopping H}, the dynamics of the physical system is highly
non-trivial, not least on account of the non-linear transformation between degrees of freedom of the physical~\eqref{eq: Hamiltonian} and dual Hamiltonian~\eqref{eq: hopping H}.

The identification of the set of conserved quantities $\left\{ q_j \right\}$, and the derivation of equation~\eqref{eq: hopping H} proceeds by a duality mapping~\cite{Kramers1941,Fradkin1978} from bond-spins $\sigma$ to site-spins $\tau$,
\begin{equation}
\hat{\tau}^z_{j}= \hat{\sigma}^x_{j-1, j} \hat{\sigma}^x_{j, j+1}, \qquad \hat{\sigma}^z_{ j, j+1} = \hat{\tau}^x_j\hat{\tau}^x_{j+1} \ .
\end{equation}
We consider a system of $N$ sites with open boundary conditions, see Fig.~\ref{fig: model}. Periodic boundary conditions introduce only a few technical differences, such as the global constraint on spins, which is automatically satisfied by our choice of initial spin-states (for more details see, e.g., Refs.~\cite{Essler2016,Iorgov2011}). In terms of the dual spins, the Hamiltonian assumes the following form
\begin{equation}\label{eq: tau H}
\hat{H} = -J\sum_{\la ij\ra} \hat{\tau}^x_i\hat{\tau}^x_{j} \hat{f}^\dag_{i} \hat{f}_{j} - h  \sum_{i} \hat{\tau}^z_i.
\end{equation}
Here $N$ mutually commuting conserved charges are given by $\hat{q}_j \equiv \hat{\tau}^z_j(-1)^{\hat{n}_j}$ with $\hat{n}_j=\hat{f}^{\dagger}_j\hat{f}_j$. The charges also commute with the Hamiltonian $\hat{H}$, but change sign under the action of operators $\hat{\tau}^x_j$, and $\hat{f}^{(\dag)}_j$. In terms of new fermion operators $\hat{c}_j = \hat{\tau}^x_j \hat{f}_j$, which commute with the charges, one arrives at the Hamiltonian~\eqref{eq: hopping H}.

We restrict initial states at $t=0$ to tensor products of fermion and spin states $| 0 \ra = | S \ra \otimes |\psi\ra$. The z-polarized initial state of bond-spins $|S\ra = |\uparrow \uparrow \uparrow \cdots \ra_\sigma$ implies a sum over all $2^N$ charge configurations $\{q_i\}=\pm1$, 
\begin{equation}\label{spin_state}
|0 \ra = \frac{1}{2^{N/2}} \sum_{\{q_i\}=\pm1} |q_1, q_2, \ldots, q_N\ra \otimes|\psi \ra, 
\end{equation}
which leads to correlators averaged over a binary potential. This particular initial spin state ensures that the tensor product form is retained after the transformation, which would not generally be the case. However, we find that this is not crucial for the observed phenomenology, see Supplementary Material for more details.
Note that states akin to (\ref{spin_state}) were suggested in \cite{Paredes2005} for quantum simulations of disordered systems. In our model this state appears naturally from a translationally invariant initial state.

\begin{figure*}[tb!]
	\centering
	\subfigimg[width=.335\textwidth,valign=t]{\hspace*{10pt} \textbf{(a) memory of the initial state}}{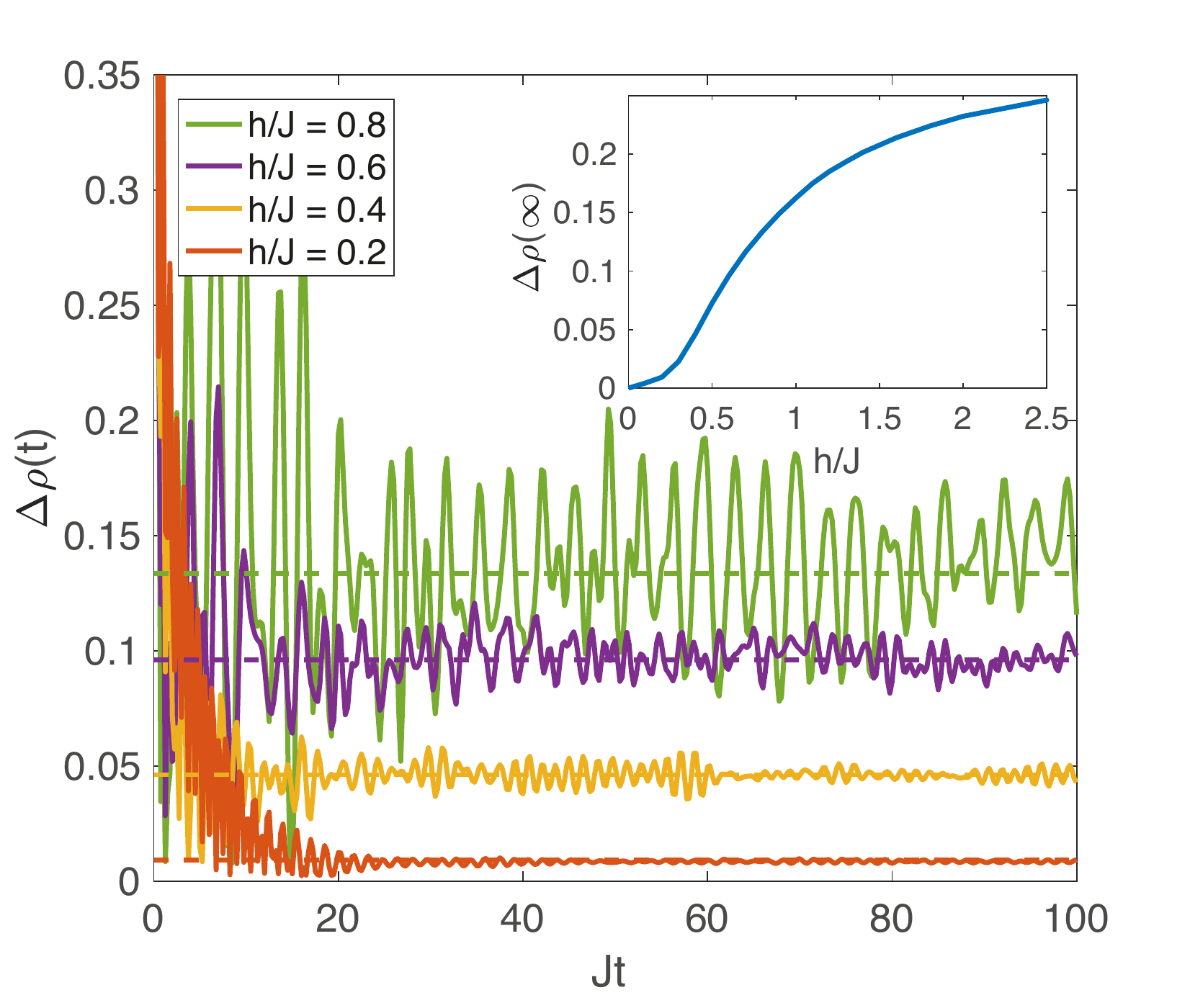}
	\!\!\!\!\!\!\!\!\subfigimg[width=.335\textwidth,valign=t]{\hspace*{10pt} \textbf{(b) domain-wall melting}}{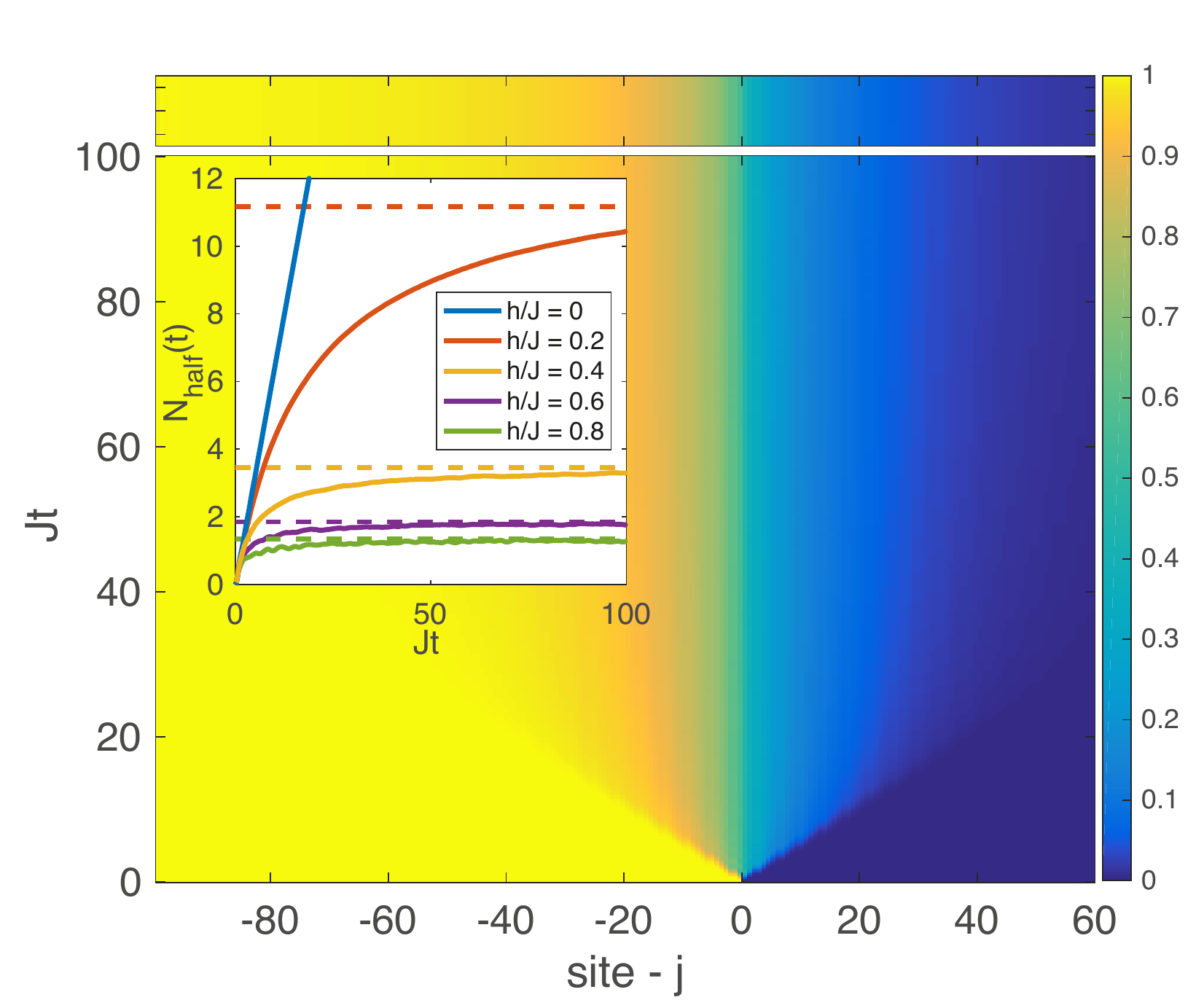}
	\subfigimg[width=.335\textwidth,valign=t]{\hspace*{10pt} \textbf{(c) correlations light-cone}}{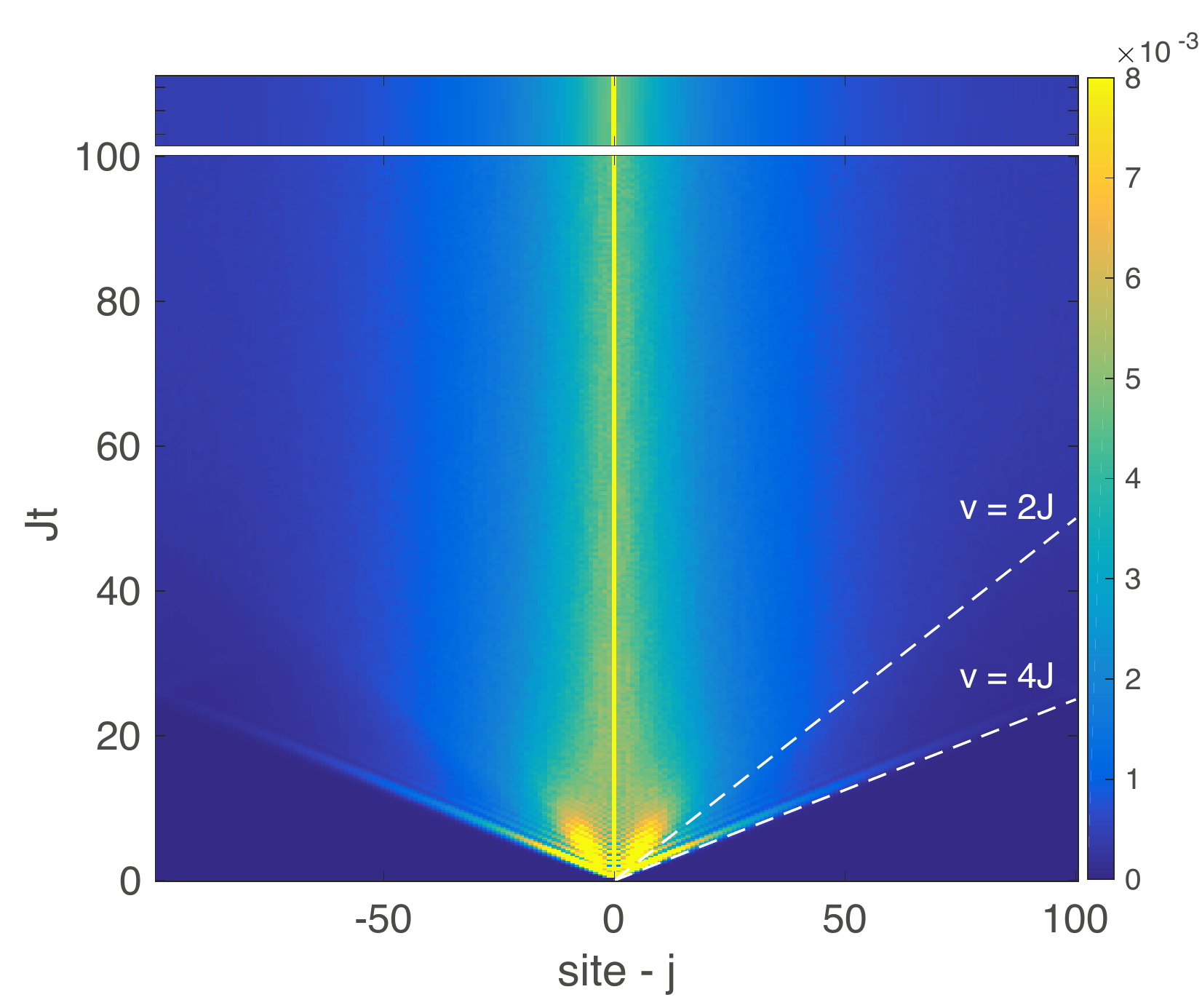}
	\caption{Time evolution of the fermion subsystem. (a) $\Delta\rho(t)$, after a quench from a density wave initial state, for a range of values $h/J$, with dashed lines showing a long-time limit; (inset) long time limit of $\Delta\rho$ as a function of $h/J$. (b) fermion density for a domain-wall initial state at $h/J = 0.3$; (inset) integrated fermion number in the right-half of the chain as a function of time. (c) absolute value of the connected density-density correlator $\la 0 | \hat{n}_l(t) \hat{n}_{l+j}(t) | 0 \ra_c$ for a density-wave initial state at $h/J = 0.2$. Dashed lines correspond to two light-cone velocities. The upper panels in (b) and (c) show the long-time limit $Jt=10^9$. All figures are computed for systems with $N=200$ sites.}\label{fig: fermions}
\end{figure*}

\textit{Results.}
The identification of the conserved charges, and the form of the dual Hamiltonian (\ref{eq: hopping H}), allows us to evaluate correlators, which we will use to demonstrate disorder-free dynamical localization. The results presented below were obtained for systems with up to $N=200$ sites, see Supplementary Material.  

\textit{Fermionic subsystem.} First, we consider the fermionic subsystem, see Fig.~\ref{fig: fermions}. For a density-wave initial state%
\begin{equation}
\Delta\rho(t) = \frac{1}{N} \sum_j |\la 0| \hat{n}_{j}(t) - \hat{n}_{j+1}(t)|0\ra|,
\end{equation}
measures the average staggered fermion density. In an ergodic phase, $\Delta\rho(t)$ vanishes at long times. In our model, it instead shows saturation to a finite asymptotic value, $\Delta\rho(\infty)$, Fig~\ref{fig: fermions}(a), which
grows monotonically with
 $h/J$ (inset). This 
demonstrates persistence of memory of the initial state. 

\begin{figure}[b!]
	\centering
	\includegraphics[width=.85\columnwidth]{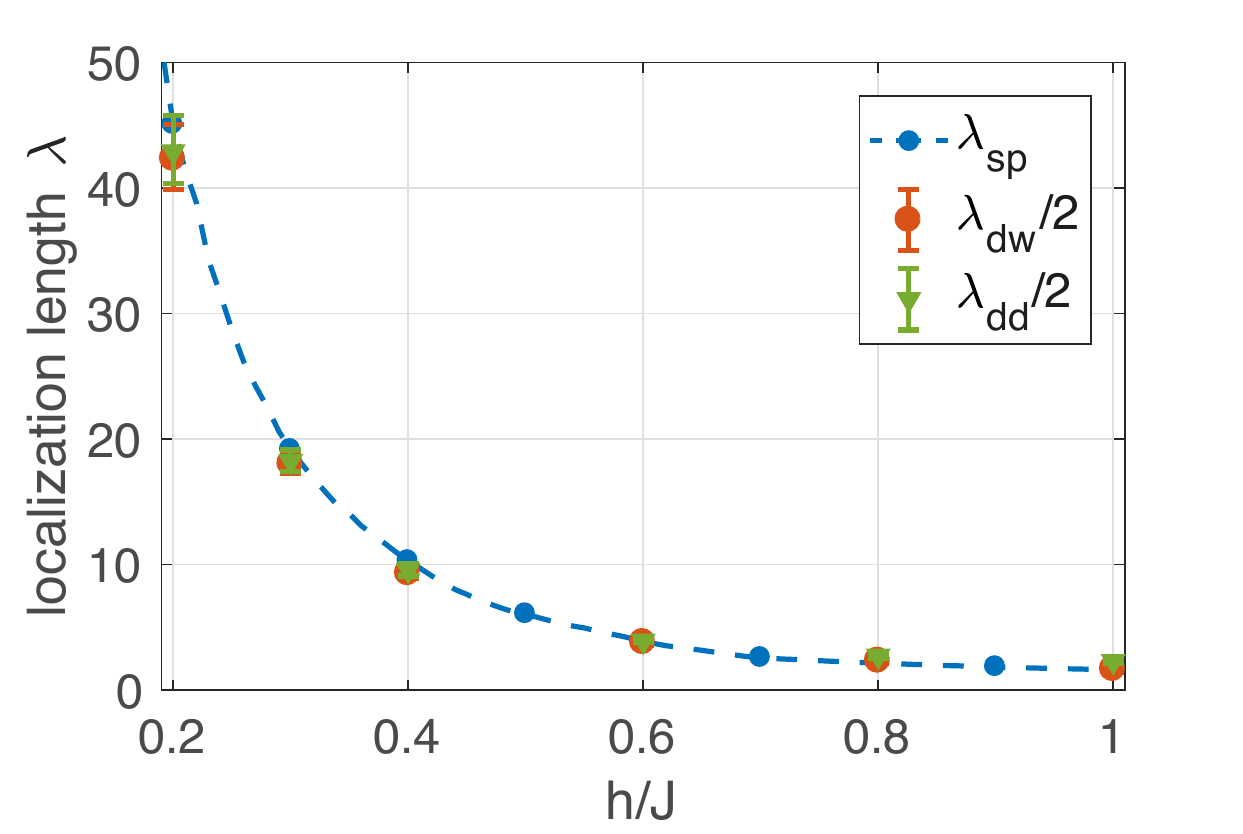}
	\caption{The localization length. Lengthscales determined from the tails $\sim\exp(-j/\lambda)$ in the long-time limit of the density imbalance ($\lambda_{dw}$ -- circles), the density-density correlators ($\lambda_{dd}$ -- triangles), and the single-particle localization length ($\lambda_{sp}$)~\cite{Kramer1993}. The error bars are given by 2.5 standard deviations of the numerical exponential fit. For $h/J=0.2,0.3$ we used $N=400$, with $N=200$ for all others.}\label{fig: localization length}
\end{figure}

Similarly, for the domain-wall initial state~\cite{Choi2016,Hauschild2016}, with a maximal
density imbalance between two halves of the lattice, Fig.~\ref{fig: fermions}(b), an initial spreading 
of fermions {\it eventually halts}, and the number of particles emitted from the filled to the empty half
of the system  
after the quench, $N_\text{half}$, (inset) remains finite. The 
long-time spatial density distribution shows exponential tails. The decay length is simply 
proportional to the single-particle localization length~\cite{Kramer1993}, as in \cite{Hauschild2016}, see Fig.~\ref{fig: localization length}, with a proportionality constant of approximately two. 

Next, we diagnose localization via connected density-density correlators $\la 0| \hat{n}_j(t) \hat{n}_k(t) | 0 \ra_c$. In the absence of dynamical disorder, $h=0$, we observe the light-cone spreading of a free-fermion model~\cite{Essler2016}, whose envelop is set by the velocity corresponding to the Lieb-Robinson bound $v_{LR}=4J$, twice the maximum fermion group-velocity. This is in contrast with quenches for $h\neq 0$ (Fig.~\ref{fig: fermions}(c)). For a density-wave initial state at short-times, the linear spreading of correlators, bounded by $v_{LR}$, and accompanied by a second signal at $v_{LR}/2$, is eventually suppressed at long times. In this limit the correlators assume a stationary form~\cite{Znidaric2008}, decaying with the same exponent as the density imbalance, Fig.~\ref{fig: localization length}. We emphasise that we find a similar localization behaviour for a translationally invariant Fermi-sea initial state, some results for which are shown in the Supplementary Material. 

This above ensemble  of  results provides unambiguous evidence of localization of the fermionic subsystem in a 
model without quenched disorder. This is our first central result.

\begin{figure}[tb!]
\centering
\subfigimg[width=.42\textwidth,valign=t]{\hspace*{10pt} \textbf{(a) spin average}}{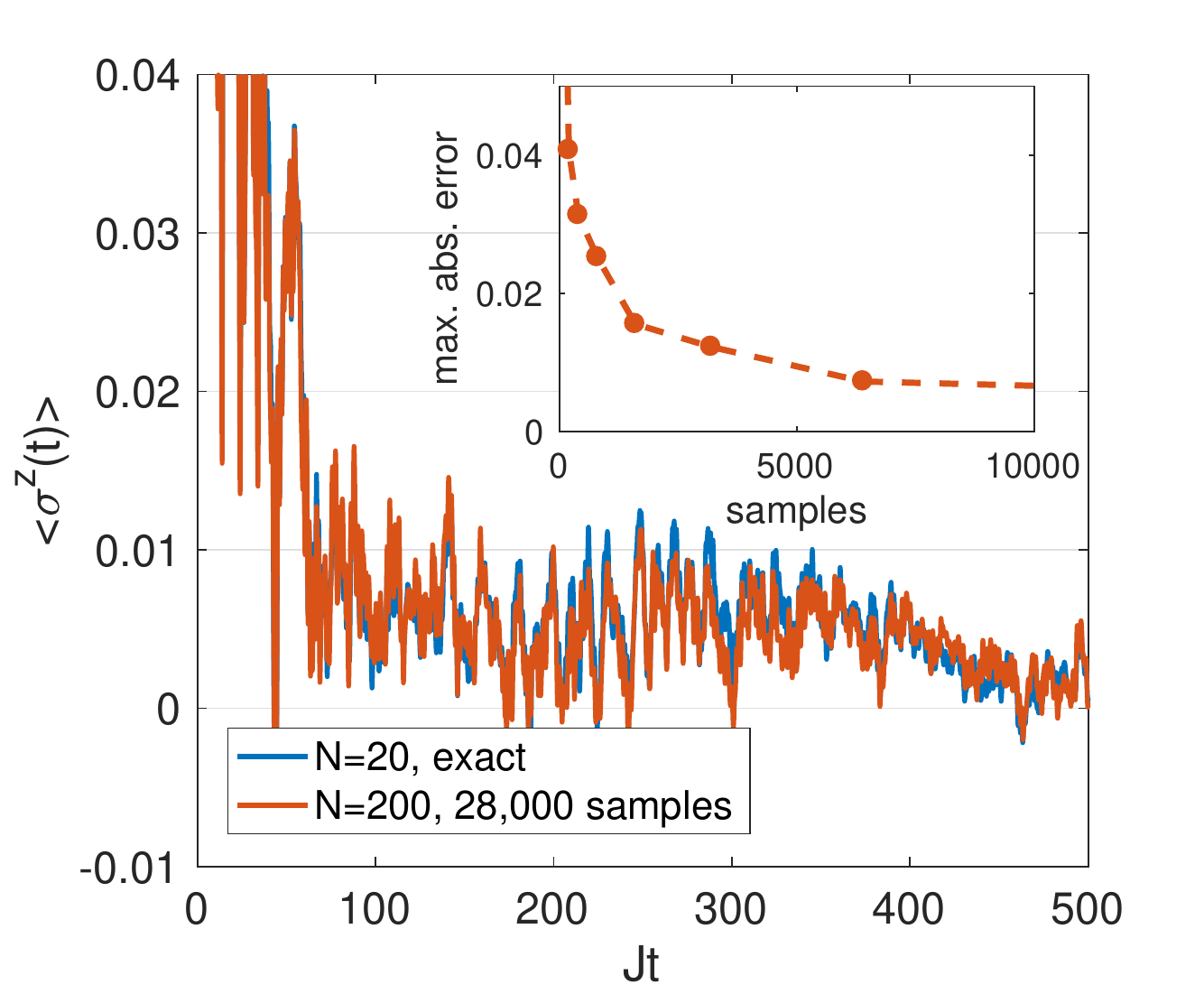}\\
\quad\subfigimg[width=.42\textwidth,valign=t]{\hspace*{7pt} \textbf{(b) spin-spin correlator}}{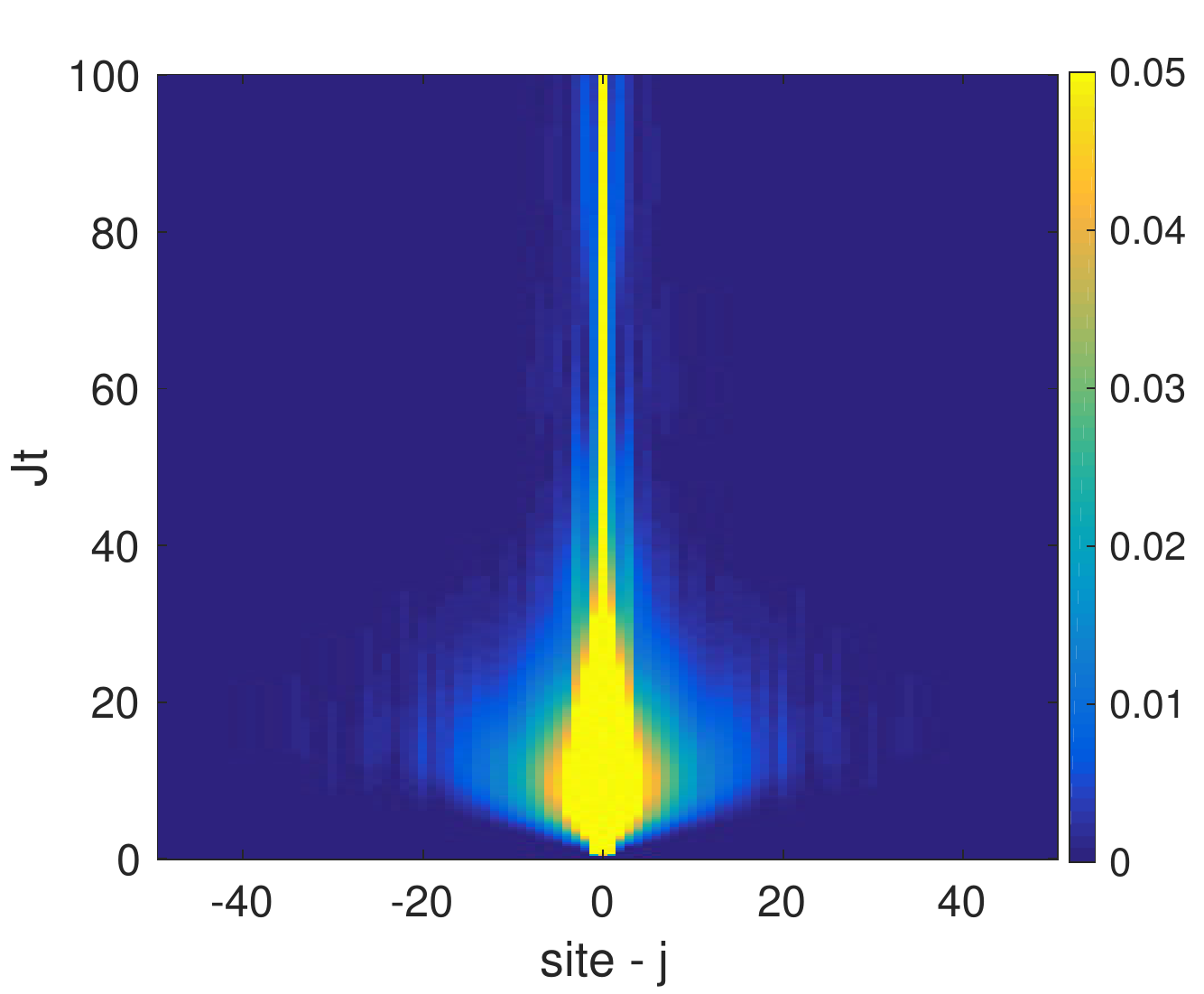}
\caption{Time evolution of the spin subsystem. (a) The spin average $\la\hat{\sigma}^z(t)\ra$ of the bond-spin at $h/J=1$ after a quench from an initial half-filled Fermi-sea state, comparing exact result for $N=20$ with disorder averaged result for $N=200$. (inset) maximum absolute error of disorder sampling, compared with exact summation over all spin configurations, as a function of number of samples for $N=15$, $h/J=1$. 
(b) absolute value of the connected spin-correlator $\la \hat{\sigma}^z_l(t) \hat{\sigma}^z_{l+j}(t) \ra_c$ for $h/J=0.3, N=100$.
}\label{fig: spin}
\end{figure}

\textit{Spin subsystem.} Let us now turn to the discussion of results for the spin subsystem. The expectation value of the $z$-component of the bond-spin, Fig.~\ref{fig: spin}(a), decays to zero at long times for all $h\neq 0$. Furthermore, for the explored range of parameters $h/J$, we find that this decay is asymptotically a power-law. The remarkable qualitative agreement between the exact result for $N=20$, and the disorder averaged result for $N=200$ suggests that the spin-dynamics is dominated by regions of  finite  size, presumably  of the order of fermion localization length. Intriguingly, we find persistent spin-fluctuations accompanying the power-law decay.

The equal-time spin-correlator on two bonds, Fig.~\ref{fig: spin}(b) exhibits an initial linear light-cone. As with the fermion correlators, the extent of the ballistic regime is determined by the localization length.
As for the spin average, we find a decay of all spatial correlations to zero in the long-time limit, 
indicating equilibration of the spin subsystem.%

\emph{Discussion.} We have observed dynamical localization after a completely translationally invariant quantum quench. The fermionic subsystem retains memory of the initial state, whereas the spin subsystem eventually equilibrates. In our model of fermions coupled to dynamical spins, the requisite randomness is generated dynamically. The main technical advance of our work is the identification of an extensive set of conserved  $\mathbb{Z}_2$ charges such that the time evolution can be described by a non-interacting Hamiltonian with effective binary disorder, allowing  numerical computations on large systems. 

Despite the close relation of our model to the heavy-light mixtures
studied in the context of quasi-MBL~\cite{Schiulaz2015,Yao2014}, we 
identify several key differences.
First, the only limit where true nonergodicity is known in these models is an infinite mass ``heavy" species, whereas the corresponding limit for us ($h\rightarrow 0$) amounts to free fermions. Otherwise the dynamics of
the heavy species generally leads to the eventual return of
ergodicity~\cite{Yao2014,Papic2015}, whereas we observe complete
localization for all $h\neq0$.
We can also vary the parameter $h/J$ freely, thus there is no meaningful
distinction in our model in terms of heavy/light particles: the two
subsystems are instead characterized by equilibration or lack thereof.
This is similar to the distinction between the components of QDL~\cite{Grover2014,Veness2016,Garrison2016}. Interestingly, the model of Eq.~\eqref{eq: hopping H} is also related to Falikov-Kimbell model Ref.~\cite{Antipov2016}, which shows a rich phase diagram at finite temperature.

The tunability of $h/J$ may be important for experimental realisations, as
varying $h/J$ can change the localization length, e.g., in the range
$0.2 \ldots 1$ from 100 down to almost a single lattice spacing. This
could enable quantum simulations even on the currently available
relatively small systems~\cite{Monroe2016}.

Moreover, our setup itself is remarkably simple: the Hamiltonian contains just nearest-neighbour exchange and hopping terms; while the initial state can be chosen as simple, entirely unentangled product state of spins and fermions, thereby removing obstacles related to preparing complex many-body states. This should enable our proposal to take maximal advantage from recent progress in quantum simulations of lattice gauge theories~\cite{Martinez2016}. 

Indeed, there are numerous connections to gauge theories appearing in other contexts. Our model can be thought of as a fermionic matter field minimally coupled to a dynamical gauge field with a somewhat unusual Hamiltonian. It is less constrained, but related to $\mathbb{Z}_2$-slave-spin representations of the Hubbard model, and of lattice gauge theories~\cite{Ruegg2010,Zitko2015,Gazit2016}. Crucially, our model allows for straightforward generalisations, in particular to higher dimensions, yielding, e.g., Kitaev's toric code model coupled to fermionic matter. This holds the promise of studying, in a broad range of settings, the novel  localization phenomena uncovered in this work.

\emph{Acknowledgements.} We are grateful to  J. T.~Chalker, F. H. L.~Essler, C.~Castelnovo, and A.~Nahum for enlightening discussions. A. S.~would like to acknowledge the EPSRC for studentship funding under Grant No.~EP/M508007/1. J. K. is supported by the Marie Curie Programme under EC Grant agreements No.703697. The work of  D. L. K.~was supported by EPSRC Grant No.~EP/M007928/1.  R. M.\ was in part supported by DFG under grant SFB 1143. 


\begin{thebibliography}{33}%
	\makeatletter
	\providecommand \@ifxundefined [1]{%
		\@ifx{#1\undefined}
	}%
	\providecommand \@ifnum [1]{%
		\ifnum #1\expandafter \@firstoftwo
		\else \expandafter \@secondoftwo
		\fi
	}%
	\providecommand \@ifx [1]{%
		\ifx #1\expandafter \@firstoftwo
		\else \expandafter \@secondoftwo
		\fi
	}%
	\providecommand \natexlab [1]{#1}%
	\providecommand \enquote  [1]{``#1''}%
	\providecommand \bibnamefont  [1]{#1}%
	\providecommand \bibfnamefont [1]{#1}%
	\providecommand \citenamefont [1]{#1}%
	\providecommand \href@noop [0]{\@secondoftwo}%
	\providecommand \href [0]{\begingroup \@sanitize@url \@href}%
	\providecommand \@href[1]{\@@startlink{#1}\@@href}%
	\providecommand \@@href[1]{\endgroup#1\@@endlink}%
	\providecommand \@sanitize@url [0]{\catcode `\\12\catcode `\$12\catcode
		`\&12\catcode `\#12\catcode `\^12\catcode `\_12\catcode `\%12\relax}%
	\providecommand \@@startlink[1]{}%
	\providecommand \@@endlink[0]{}%
	\providecommand \url  [0]{\begingroup\@sanitize@url \@url }%
	\providecommand \@url [1]{\endgroup\@href {#1}{\urlprefix }}%
	\providecommand \urlprefix  [0]{URL }%
	\providecommand \Eprint [0]{\href }%
	\providecommand \doibase [0]{http://dx.doi.org/}%
	\providecommand \selectlanguage [0]{\@gobble}%
	\providecommand \bibinfo  [0]{\@secondoftwo}%
	\providecommand \bibfield  [0]{\@secondoftwo}%
	\providecommand \translation [1]{[#1]}%
	\providecommand \BibitemOpen [0]{}%
	\providecommand \bibitemStop [0]{}%
	\providecommand \bibitemNoStop [0]{.\EOS\space}%
	\providecommand \EOS [0]{\spacefactor3000\relax}%
	\providecommand \BibitemShut  [1]{\csname bibitem#1\endcsname}%
	\let\auto@bib@innerbib\@empty
	\bibitem [{\citenamefont {Anderson}(1958)}]{Anderson1958}%
	\BibitemOpen
	\bibfield  {author} {\bibinfo {author} {\bibfnamefont {P.~W.}\ \bibnamefont
			{Anderson}},\ }\href {\doibase 10.1103/PhysRev.109.1492} {\bibfield
		{journal} {\bibinfo  {journal} {Phys. Rev.}\ }\textbf {\bibinfo {volume}
			{109}},\ \bibinfo {pages} {1492} (\bibinfo {year} {1958})}\BibitemShut
	{NoStop}%
	\bibitem [{\citenamefont {Basko}\ \emph {et~al.}(2006)\citenamefont {Basko},
		\citenamefont {Aleiner},\ and\ \citenamefont {Altshuler}}]{Basko2006}%
	\BibitemOpen
	\bibfield  {author} {\bibinfo {author} {\bibfnamefont {D.}~\bibnamefont
			{Basko}}, \bibinfo {author} {\bibfnamefont {I.}~\bibnamefont {Aleiner}}, \
		and\ \bibinfo {author} {\bibfnamefont {B.}~\bibnamefont {Altshuler}},\ }\href
	{\doibase 10.1016/j.aop.2005.11.014} {\bibfield  {journal} {\bibinfo
			{journal} {Ann. Phys. (N. Y).}\ }\textbf {\bibinfo {volume} {321}},\ \bibinfo
		{pages} {1126} (\bibinfo {year} {2006})}\BibitemShut {NoStop}%
	\bibitem [{\citenamefont {Vasseur}\ and\ \citenamefont
		{Moore}(2016)}]{Vasseur2016}%
	\BibitemOpen
	\bibfield  {author} {\bibinfo {author} {\bibfnamefont {R.}~\bibnamefont
			{Vasseur}}\ and\ \bibinfo {author} {\bibfnamefont {J.~E.}\ \bibnamefont
			{Moore}},\ }\href {\doibase 10.1088/1742-5468/2016/06/064010} {\bibfield
		{journal} {\bibinfo  {journal} {J. Stat. Mech. Theory Exp.}\ }\textbf
		{\bibinfo {volume} {2016}},\ \bibinfo {pages} {064010} (\bibinfo {year}
		{2016})}\BibitemShut {NoStop}%
	\bibitem [{\citenamefont {Nandkishore}\ and\ \citenamefont
		{Huse}(2015)}]{Nandkishore2015}%
	\BibitemOpen
	\bibfield  {author} {\bibinfo {author} {\bibfnamefont {R.}~\bibnamefont
			{Nandkishore}}\ and\ \bibinfo {author} {\bibfnamefont {D.~A.}\ \bibnamefont
			{Huse}},\ }\href {\doibase 10.1146/annurev-conmatphys-031214-014726}
	{\bibfield  {journal} {\bibinfo  {journal} {Annu. Rev. Condens. Matter
				Phys.}\ }\textbf {\bibinfo {volume} {6}},\ \bibinfo {pages} {15} (\bibinfo
		{year} {2015})}\BibitemShut {NoStop}%
	\bibitem [{\citenamefont {Kagan}\ and\ \citenamefont {Maksimov}(1984)}]{Kagan}%
	\BibitemOpen
	\bibfield  {author} {\bibinfo {author} {\bibfnamefont {Y.}~\bibnamefont
			{Kagan}}\ and\ \bibinfo {author} {\bibfnamefont {L.~A.}\ \bibnamefont
			{Maksimov}},\ }\href {\doibase 0038-5646/84/070201-10} {\bibfield  {journal}
		{\bibinfo  {journal} {Sov. Phys. JETP}\ }\textbf {\bibinfo {volume} {60}},\ \bibinfo
		{pages} {201} (\bibinfo {year} {1984})}\BibitemShut {NoStop}%
	\bibitem [{\citenamefont {Yao}\ \emph {et~al.}(2016)\citenamefont {Yao},
		\citenamefont {Laumann}, \citenamefont {Cirac}, \citenamefont {Lukin},\ and\
		\citenamefont {Moore}}]{Yao2014}%
	\BibitemOpen
	\bibfield  {author} {\bibinfo {author} {\bibfnamefont {N.~Y.}\ \bibnamefont
			{Yao}}, \bibinfo {author} {\bibfnamefont {C.~R.}\ \bibnamefont {Laumann}},
		\bibinfo {author} {\bibfnamefont {J.~I.}\ \bibnamefont {Cirac}}, \bibinfo
		{author} {\bibfnamefont {M.~D.}\ \bibnamefont {Lukin}}, \ and\ \bibinfo
		{author} {\bibfnamefont {J.~E.}\ \bibnamefont {Moore}},\ }\href {\doibase
		10.1103/PhysRevLett.117.240601} {\bibfield  {journal} {\bibinfo  {journal}
			{Phys. Rev. Lett.}\ }\textbf {\bibinfo {volume} {117}},\ \bibinfo {pages}
		{240601} (\bibinfo {year} {2016})}\BibitemShut {NoStop}%
	\bibitem [{\citenamefont {Schiulaz}\ \emph {et~al.}(2015)\citenamefont
		{Schiulaz}, \citenamefont {Silva},\ and\ \citenamefont
		{M\"{u}ller}}]{Schiulaz2015}%
	\BibitemOpen
	\bibfield  {author} {\bibinfo {author} {\bibfnamefont {M.}~\bibnamefont
			{Schiulaz}}, \bibinfo {author} {\bibfnamefont {A.}~\bibnamefont {Silva}}, \
		and\ \bibinfo {author} {\bibfnamefont {M.}~\bibnamefont {M\"{u}ller}},\
	}\href {\doibase 10.1103/PhysRevB.91.184202} {\bibfield  {journal} {\bibinfo
		{journal} {Phys. Rev. B}\ }\textbf {\bibinfo {volume} {91}},\ \bibinfo
	{pages} {184202} (\bibinfo {year} {2015})}\BibitemShut {NoStop}%
\bibitem [{\citenamefont {Enss}\ \emph {et~al.}(2016)\citenamefont {Enss},
	\citenamefont {Andraschko},\ and\ \citenamefont {Sirker}}]{Enss2016}%
\BibitemOpen
\bibfield  {author} {\bibinfo {author} {\bibfnamefont {T.}~\bibnamefont
		{Enss}}, \bibinfo {author} {\bibfnamefont {F.}~\bibnamefont {Andraschko}}, \
	and\ \bibinfo {author} {\bibfnamefont {J.}~\bibnamefont {Sirker}},\ }\href {\doibase 10.1103/PhysRevB.95.045121} {\bibfield  {journal} {\bibinfo  {journal}
		{Phys. Rev. B}\ }\textbf {\bibinfo {volume} {95}},\ \bibinfo
	{pages} {045121} (\bibinfo {year} {2017})}\BibitemShut {NoStop}%
\bibitem []{Schreiber2015}%
\BibitemOpen
\bibfield  {author} {\bibinfo {author} {\bibfnamefont {M.}\ \bibnamefont
		{Schreiber}}, \bibinfo {author} {\bibfnamefont {S.~S.}~\bibnamefont {Hodgman}},
	\bibinfo {author} {\bibfnamefont {S.}\ \bibnamefont {Bordia}}, \bibinfo
	{author} {\bibfnamefont {H.~P.}~\bibnamefont {L{\"u}schen}}, \bibinfo {author}
	{\bibfnamefont {M.~H.}~\bibnamefont {Fischer}}, \bibinfo {author} {\bibfnamefont
		{R.}\ \bibnamefont {Vosk}}, \bibinfo {author} {\bibfnamefont {E.}~\bibnamefont
		{Altman}}, \bibinfo {author} {\bibfnamefont {U.}~\bibnamefont {Schneider}}, \
	and\ \bibinfo {author} {\bibfnamefont {I.}~\bibnamefont {Bloch}},\ }\href {\doibase 10.1126/science.aaa7432} {\bibfield  {journal}
	{\bibinfo  {journal} {Science}\ }\textbf {\bibinfo {volume} {349}},\ \bibinfo
	{pages} {842} (\bibinfo {year} {2015})}\BibitemShut {NoStop}%
\bibitem [{\citenamefont {Choi}\ \emph {et~al.}(2016)\citenamefont {Choi},
	\citenamefont {Hild}, \citenamefont {Zeiher}, \citenamefont {Schauss},
	\citenamefont {Rubio-Abadal}, \citenamefont {Yefsah}, \citenamefont
	{Khemani}, \citenamefont {Huse}, \citenamefont {Bloch},\ and\ \citenamefont
	{Gross}}]{Choi2016}%
\BibitemOpen
\bibfield  {author} {\bibinfo {author} {\bibfnamefont {J.-Y.}\ \bibnamefont
		{Choi}}, \bibinfo {author} {\bibfnamefont {S.}~\bibnamefont {Hild}}, \bibinfo
	{author} {\bibfnamefont {J.}~\bibnamefont {Zeiher}}, \bibinfo {author}
	{\bibfnamefont {P.}~\bibnamefont {Schauss}}, \bibinfo {author} {\bibfnamefont
		{A.}~\bibnamefont {Rubio-Abadal}}, \bibinfo {author} {\bibfnamefont
		{T.}~\bibnamefont {Yefsah}}, \bibinfo {author} {\bibfnamefont
		{V.}~\bibnamefont {Khemani}}, \bibinfo {author} {\bibfnamefont {D.~A.}\
		\bibnamefont {Huse}}, \bibinfo {author} {\bibfnamefont {I.}~\bibnamefont
		{Bloch}}, \ and\ \bibinfo {author} {\bibfnamefont {C.}~\bibnamefont
		{Gross}},\ }\href {\doibase 10.1126/science.aaf8834} {\bibfield  {journal}
	{\bibinfo  {journal} {Science}\ }\textbf {\bibinfo {volume} {352}},\ \bibinfo
	{pages} {1547} (\bibinfo {year} {2016})}\BibitemShut {NoStop}%
\bibitem [{\citenamefont {Hauschild}\ \emph {et~al.}(2016)\citenamefont
	{Hauschild}, \citenamefont {Heidrich-Meisner},\ and\ \citenamefont
	{Pollmann}}]{Hauschild2016}%
\BibitemOpen
\bibfield  {author} {\bibinfo {author} {\bibfnamefont {J.}~\bibnamefont
		{Hauschild}}, \bibinfo {author} {\bibfnamefont {F.}~\bibnamefont
		{Heidrich-Meisner}}, \ and\ \bibinfo {author} {\bibfnamefont
		{F.}~\bibnamefont {Pollmann}},\ }\href {\doibase 10.1103/PhysRevB.94.161109}
{\bibfield  {journal} {\bibinfo  {journal} {Phys. Rev. B}\ }\textbf {\bibinfo
		{volume} {94}},\ \bibinfo {pages} {161109} (\bibinfo {year}
	{2016})}\BibitemShut {NoStop}%
\bibitem [{\citenamefont {Essler}\ and\ \citenamefont
	{Fagotti}(2016)}]{Essler2016}%
\BibitemOpen
\bibfield  {author} {\bibinfo {author} {\bibfnamefont {F.~H.~L.}\
		\bibnamefont {Essler}}\ and\ \bibinfo {author} {\bibfnamefont
		{M.}~\bibnamefont {Fagotti}},\ }\href {\doibase
	10.1088/1742-5468/2016/06/064002} {\bibfield  {journal} {\bibinfo  {journal}
		{J. Stat. Mech. Theory Exp.}\ }\textbf {\bibinfo {volume} {2016}},\ \bibinfo
	{pages} {064002} (\bibinfo {year} {2016})}\BibitemShut {NoStop}%
\bibitem [{\citenamefont {\v{Z}nidari\v{c}}\ \emph {et~al.}(2008)\citenamefont
	{\v{Z}nidari\v{c}}, \citenamefont {Prosen},\ and\ \citenamefont
	{Prelov\v{s}ek}}]{Znidaric2008}%
\BibitemOpen
\bibfield  {author} {\bibinfo {author} {\bibfnamefont {M.}~\bibnamefont
		{\v{Z}nidari\v{c}}}, \bibinfo {author} {\bibfnamefont {T.}~\bibnamefont
		{Prosen}}, \ and\ \bibinfo {author} {\bibfnamefont {P.}~\bibnamefont
		{Prelov\v{s}ek}},\ }\href {\doibase 10.1103/PhysRevB.77.064426} {\bibfield
	{journal} {\bibinfo  {journal} {Phys. Rev. B}\ }\textbf {\bibinfo {volume}
		{77}},\ \bibinfo {pages} {064426} (\bibinfo {year} {2008})}\BibitemShut
{NoStop}%
\bibitem [{\citenamefont {Papi\'{c}}\ \emph {et~al.}(2015)\citenamefont
	{Papi\'{c}}, \citenamefont {Stoudenmire},\ and\ \citenamefont
	{Abanin}}]{Papic2015}%
\BibitemOpen
\bibfield  {author} {\bibinfo {author} {\bibfnamefont {Z.}~\bibnamefont
		{Papi\'{c}}}, \bibinfo {author} {\bibfnamefont {E.~M.}\ \bibnamefont
		{Stoudenmire}}, \ and\ \bibinfo {author} {\bibfnamefont {D.~A.}\ \bibnamefont
		{Abanin}},\ }\href {\doibase 10.1016/j.aop.2015.08.024} {\bibfield  {journal}
	{\bibinfo  {journal} {Ann. Phys. (N. Y).}\ }\textbf {\bibinfo {volume}
		{362}},\ \bibinfo {pages} {714} (\bibinfo {year} {2015})}\BibitemShut
{NoStop}%
\bibitem [{\citenamefont {van Horssen}\ \emph {et~al.}(2015)\citenamefont {van
		Horssen}, \citenamefont {Levi},\ and\ \citenamefont
	{Garrahan}}]{Horssen2015}%
\BibitemOpen
\bibfield  {author} {\bibinfo {author} {\bibfnamefont {M.}~\bibnamefont {van
			Horssen}}, \bibinfo {author} {\bibfnamefont {E.}~\bibnamefont {Levi}}, \ and\
	\bibinfo {author} {\bibfnamefont {J.~P.}\ \bibnamefont {Garrahan}},\ }\href
{\doibase 10.1103/PhysRevB.92.100305} {\bibfield  {journal} {\bibinfo
		{journal} {Phys. Rev. B}\ }\textbf {\bibinfo {volume} {92}},\ \bibinfo
	{pages} {100305} (\bibinfo {year} {2015})}\BibitemShut {NoStop}%
	\bibitem [{\citenamefont {Grover}\ and\ \citenamefont
		{Fisher}(2014)}]{Grover2014}%
	\BibitemOpen
	\bibfield  {author} {\bibinfo {author} {\bibfnamefont {T.}~\bibnamefont
			{Grover}}\ and\ \bibinfo {author} {\bibfnamefont {M.~P.~A.}\ \bibnamefont
			{Fisher}},\ }\href {\doibase 10.1088/1742-5468/2014/10/P10010} {\bibfield
		{journal} {\bibinfo  {journal} {J. Stat. Mech. Theory Exp.}\ }\textbf
		{\bibinfo {volume} {2014}},\ \bibinfo {pages} {P10010} (\bibinfo {year}
		{2014})}\BibitemShut {NoStop}%
\bibitem [{\citenamefont {Veness}\ \emph {et~al.}(2016)\citenamefont {Veness},
	\citenamefont {Essler},\ and\ \citenamefont {Fisher}}]{Veness2016}%
\BibitemOpen
\bibfield  {author} {\bibinfo {author} {\bibfnamefont {T.}~\bibnamefont
		{Veness}}, \bibinfo {author} {\bibfnamefont {F.~H.~L.}\ \bibnamefont
		{Essler}}, \ and\ \bibinfo {author} {\bibfnamefont {M.~P.~A.}~\bibnamefont
		{Fisher}},\ }\href {https://arxiv.org/abs/1611.02075} {\bibfield  {journal}
	{\bibinfo  {journal} {arXiv:1611.02075}\ } (\bibinfo {year}
	{2016})}\BibitemShut {NoStop}%
\bibitem [{\citenamefont {Garrison}\ \emph {et~al.}(2016)\citenamefont {Garrison},
	\citenamefont {Mishmash},\ and\ \citenamefont {Fisher}}]{Garrison2016}%
\BibitemOpen
\bibfield  {author} {\bibinfo {author} {\bibfnamefont {J.~R.}~\bibnamefont
		{Garrison}}, \bibinfo {author} {\bibfnamefont {R.~V.}\ \bibnamefont
		{Mishmash}}, \ and\ \bibinfo {author} {\bibfnamefont {M.~P.~A}~\bibnamefont
		{Fisher}},\ }\href {\doibase 10.1103/PhysRevB.95.054204} {\bibfield  {journal}
	{\bibinfo  {journal} {Phys. Rev. B}\ }\textbf {\bibinfo {volume} {95}},\
	\bibinfo {pages} {054204} (\bibinfo {year}
	{2017})}\BibitemShut {NoStop}%
\bibitem [{\citenamefont {Dunlap}\ and\ \citenamefont
	{Kenkre}(1986)}]{PhysRevB.34.3625}%
\BibitemOpen
\bibfield  {author} {\bibinfo {author} {\bibfnamefont {D.~H.}\ \bibnamefont
		{Dunlap}}\ and\ \bibinfo {author} {\bibfnamefont {V.~M.}\ \bibnamefont
		{Kenkre}},\ }\href {\doibase 10.1103/PhysRevB.34.3625} {\bibfield  {journal}
	{\bibinfo  {journal} {Phys. Rev. B}\ }\textbf {\bibinfo {volume} {34}},\
	\bibinfo {pages} {3625} (\bibinfo {year} {1986})}\BibitemShut {NoStop}%
\bibitem [{\citenamefont {Vidal}\ \emph {et~al.}(1998)\citenamefont {Vidal},
	\citenamefont {Mosseri},\ and\ \citenamefont {Dou\ifmmode~\mbox{\c{c}}\else
		\c{c}\fi{}ot}}]{PhysRevLett.81.5888}%
\BibitemOpen
\bibfield  {author} {\bibinfo {author} {\bibfnamefont {J.}~\bibnamefont
		{Vidal}}, \bibinfo {author} {\bibfnamefont {R.}~\bibnamefont {Mosseri}}, \
	and\ \bibinfo {author} {\bibfnamefont {B.}~\bibnamefont
		{Dou\ifmmode~\mbox{\c{c}}\else \c{c}\fi{}ot}},\ }\href {\doibase
	10.1103/PhysRevLett.81.5888} {\bibfield  {journal} {\bibinfo  {journal}
		{Phys. Rev. Lett.}\ }\textbf {\bibinfo {volume} {81}},\ \bibinfo {pages}
	{5888} (\bibinfo {year} {1998})}\BibitemShut {NoStop}%
\bibitem [{\citenamefont {Kramers}\ and\ \citenamefont
	{Wannier}(1941)}]{Kramers1941}%
\BibitemOpen
\bibfield  {author} {\bibinfo {author} {\bibfnamefont {H.~A.}\ \bibnamefont
		{Kramers}}\ and\ \bibinfo {author} {\bibfnamefont {G.~H.}\ \bibnamefont
		{Wannier}},\ }\href {\doibase 10.1103/PhysRev.60.252} {\bibfield  {journal}
	{\bibinfo  {journal} {Phys. Rev.}\ }\textbf {\bibinfo {volume} {60}},\
	\bibinfo {pages} {252} (\bibinfo {year} {1941})}\BibitemShut {NoStop}%
\bibitem [{\citenamefont {Fradkin}\ and\ \citenamefont
	{Susskind}(1978)}]{Fradkin1978}%
\BibitemOpen
\bibfield  {author} {\bibinfo {author} {\bibfnamefont {E.}~\bibnamefont
		{Fradkin}}\ and\ \bibinfo {author} {\bibfnamefont {L.}~\bibnamefont
		{Susskind}},\ }\href {\doibase 10.1103/PhysRevD.17.2637} {\bibfield
	{journal} {\bibinfo  {journal} {Phys. Rev. D}\ }\textbf {\bibinfo {volume}
		{17}},\ \bibinfo {pages} {2637} (\bibinfo {year} {1978})}\BibitemShut
{NoStop}%
\bibitem [{\citenamefont {Iorgov}\ \emph {et~al.}(2011)\citenamefont {Iorgov},
	\citenamefont {Shadura},\ and\ \citenamefont {Tykhyy}}]{Iorgov2011}%
\BibitemOpen
\bibfield  {author} {\bibinfo {author} {\bibfnamefont {N.}~\bibnamefont
		{Iorgov}}, \bibinfo {author} {\bibfnamefont {V.}~\bibnamefont {Shadura}}, \
	and\ \bibinfo {author} {\bibfnamefont {Y.}~\bibnamefont {Tykhyy}},\ }\href
{\doibase 10.1088/1742-5468/2011/02/P02028} {\bibfield  {journal} {\bibinfo
		{journal} {J. Stat. Mech. Theory Exp.}\ }\textbf {\bibinfo {volume} {2011}},\
	\bibinfo {pages} {P02028} (\bibinfo {year} {2011})}\BibitemShut {NoStop}%
\bibitem [{\citenamefont {Paredes}\ \emph {et~al.}(2005)\citenamefont
	{Paredes}, \citenamefont {Verstraete},\ and\ \citenamefont
	{Cirac}}]{Paredes2005}%
\BibitemOpen
\bibfield  {author} {\bibinfo {author} {\bibfnamefont {B.}~\bibnamefont
		{Paredes}}, \bibinfo {author} {\bibfnamefont {F.}~\bibnamefont {Verstraete}},
	\ and\ \bibinfo {author} {\bibfnamefont {J.~I.}\ \bibnamefont {Cirac}},\
}\href {\doibase 10.1103/PhysRevLett.95.140501} {\bibfield  {journal}
{\bibinfo  {journal} {Phys. Rev. Lett.}\ }\textbf {\bibinfo {volume} {95}},\
\bibinfo {pages} {140501} (\bibinfo {year} {2005})}\BibitemShut {NoStop}%
\bibitem [{\citenamefont {Kramer}\ and\ \citenamefont
	{MacKinnon}(1993)}]{Kramer1993}%
\BibitemOpen
\bibfield  {author} {\bibinfo {author} {\bibfnamefont {B.}~\bibnamefont
		{Kramer}}\ and\ \bibinfo {author} {\bibfnamefont {A.}~\bibnamefont
		{MacKinnon}},\ }\href {\doibase 10.1088/0034-4885/56/12/001} {\bibfield
	{journal} {\bibinfo  {journal} {Reports Prog. Phys.}\ }\textbf {\bibinfo
		{volume} {56}},\ \bibinfo {pages} {1469} (\bibinfo {year}
	{1993})}\BibitemShut {NoStop}%
\bibitem [{\citenamefont {Antipov}\ \emph {et~al.}(2005)\citenamefont
	{Antipov}, \citenamefont {Javanmard}, \citenamefont {Ribeiro},\ and\ \citenamefont
	{Kirchner}}]{Antipov2016}%
\BibitemOpen
\bibfield  {author} {\bibinfo {author} {\bibfnamefont {A.~E.}~\bibnamefont
		{Antipov}}, \bibinfo {author} {\bibfnamefont {Y.}~\bibnamefont {Javanmard}}, \bibinfo {author} {\bibfnamefont {P.}~\bibnamefont {Ribeiro}},\
	and\ \bibinfo {author} {\bibfnamefont {S.}~\bibnamefont
		{Kirchner},\ }}\href {\doibase
	10.1103/PhysRevLett.117.146601} {\bibfield  {journal} {\bibinfo  {journal}
		{Phys. Rev. Lett.}\ }\textbf {\bibinfo {volume} {117}},\ \bibinfo {pages}
	{146601} (\bibinfo {year} {2016})}\BibitemShut {NoStop}%
\bibitem [{\citenamefont {Zhang}\ \emph {et~al.}(2016)\citenamefont {Zhang},
	\citenamefont {Hess}, \citenamefont {Kyprianidis}, \citenamefont {Becker},
	\citenamefont {Lee}, \citenamefont {Smith}, \citenamefont {Pagano},
	\citenamefont {Potirniche}, \citenamefont {Potter}, \citenamefont
	{Vishwanath}, \citenamefont {Yao},\ and\ \citenamefont
	{Monroe}}]{Monroe2016}%
\BibitemOpen
\bibfield  {author} {\bibinfo {author} {\bibfnamefont {J.}~\bibnamefont
		{Zhang}}, \bibinfo {author} {\bibfnamefont {P.~W.}\ \bibnamefont {Hess}},
	\bibinfo {author} {\bibfnamefont {A.}~\bibnamefont {Kyprianidis}}, \bibinfo
	{author} {\bibfnamefont {P.}~\bibnamefont {Becker}}, \bibinfo {author}
	{\bibfnamefont {A.}~\bibnamefont {Lee}}, \bibinfo {author} {\bibfnamefont
		{J.}~\bibnamefont {Smith}}, \bibinfo {author} {\bibfnamefont
		{G.}~\bibnamefont {Pagano}}, \bibinfo {author} {\bibfnamefont {I.~D.}\
		\bibnamefont {Potirniche}}, \bibinfo {author} {\bibfnamefont {A.~C.}\
		\bibnamefont {Potter}}, \bibinfo {author} {\bibfnamefont {A.}~\bibnamefont
		{Vishwanath}}, \bibinfo {author} {\bibfnamefont {N.~Y.}\ \bibnamefont {Yao}},
	\ and\ \bibinfo {author} {\bibfnamefont {C.}~\bibnamefont {Monroe}},\ } \href {\doibase
	10.1038/nature21413}{\bibfield  {journal}
	{\bibinfo  {journal} {Nature (London)}\ }  \textbf {\bibinfo {volume} {534}},\ \bibinfo {pages}{516} (\bibinfo {year}
	{2017})}\BibitemShut {NoStop}%
\bibitem [{\citenamefont {Martinez}\ \emph {et~al.}(2016)\citenamefont
	{Martinez}, \citenamefont {Muschik}, \citenamefont {Schindler}, \citenamefont
	{Nigg}, \citenamefont {Erhard}, \citenamefont {Heyl}, \citenamefont {Hauke},
	\citenamefont {Dalmonte}, \citenamefont {Monz}, \citenamefont {Zoller},\ and\
	\citenamefont {Blatt}}]{Martinez2016}%
\BibitemOpen
\bibfield  {author} {\bibinfo {author} {\bibfnamefont {E.~A.}\ \bibnamefont
		{Martinez}}, \bibinfo {author} {\bibfnamefont {C.~A.}\ \bibnamefont
		{Muschik}}, \bibinfo {author} {\bibfnamefont {P.}~\bibnamefont {Schindler}},
	\bibinfo {author} {\bibfnamefont {D.}~\bibnamefont {Nigg}}, \bibinfo {author}
	{\bibfnamefont {A.}~\bibnamefont {Erhard}}, \bibinfo {author} {\bibfnamefont
		{M.}~\bibnamefont {Heyl}}, \bibinfo {author} {\bibfnamefont {P.}~\bibnamefont
		{Hauke}}, \bibinfo {author} {\bibfnamefont {M.}~\bibnamefont {Dalmonte}},
	\bibinfo {author} {\bibfnamefont {T.}~\bibnamefont {Monz}}, \bibinfo {author}
	{\bibfnamefont {P.}~\bibnamefont {Zoller}}, \ and\ \bibinfo {author}
	{\bibfnamefont {R.}~\bibnamefont {Blatt}},\ }\href {\doibase
	10.1038/nature18318} {\bibfield  {journal} {\bibinfo  {journal} {Nature}\
	}\textbf {\bibinfo {volume} {534}},\ \bibinfo {pages} {516} (\bibinfo {year}
	{2016})}\BibitemShut {NoStop}%
\bibitem [{\citenamefont {R\"{u}egg}\ \emph {et~al.}(2010)\citenamefont
	{R\"{u}egg}, \citenamefont {Huber},\ and\ \citenamefont
	{Sigrist}}]{Ruegg2010}%
\BibitemOpen
\bibfield  {author} {\bibinfo {author} {\bibfnamefont {A.}~\bibnamefont
		{R\"{u}egg}}, \bibinfo {author} {\bibfnamefont {S.~D.}\ \bibnamefont
		{Huber}}, \ and\ \bibinfo {author} {\bibfnamefont {M.}~\bibnamefont
		{Sigrist}},\ }\href {\doibase 10.1103/PhysRevB.81.155118} {\bibfield
	{journal} {\bibinfo  {journal} {Phys. Rev. B}\ }\textbf {\bibinfo {volume}
		{81}},\ \bibinfo {pages} {155118} (\bibinfo {year} {2010})}\BibitemShut
{NoStop}%
\bibitem [{\citenamefont {\v{Z}itko}\ and\ \citenamefont
	{Fabrizio}(2015)}]{Zitko2015}%
\BibitemOpen
\bibfield  {author} {\bibinfo {author} {\bibfnamefont {R.}~\bibnamefont
		{\v{Z}itko}}\ and\ \bibinfo {author} {\bibfnamefont {M.}~\bibnamefont
		{Fabrizio}},\ }\href {\doibase 10.1103/PhysRevB.91.245130} {\bibfield
	{journal} {\bibinfo  {journal} {Phys. Rev. B}\ }\textbf {\bibinfo {volume}
		{91}},\ \bibinfo {pages} {245130} (\bibinfo {year} {2015})}\BibitemShut
{NoStop}%
\bibitem [{\citenamefont {Gazit}\ \emph {et~al.}(2016)\citenamefont {Gazit},
	\citenamefont {Randeria},\ and\ \citenamefont {Vishwanath}}]{Gazit2016}%
\BibitemOpen
\bibfield  {author} {\bibinfo {author} {\bibfnamefont {S.}~\bibnamefont
		{Gazit}}, \bibinfo {author} {\bibfnamefont {M.}~\bibnamefont {Randeria}}, \
	and\ \bibinfo {author} {\bibfnamefont {A.}~\bibnamefont {Vishwanath}},\
}\href {\doibase 10.1038/nphys4028} {\bibfield  {journal} {\bibinfo
	{journal} {Nat. Phys.}\ }\textbf {\bibinfo {volume}
	{13}},\ \bibinfo {pages} {484} (\bibinfo {year} {2017})}\BibitemShut
{NoStop}%
\end{thebibliography}

\begin{thebibliography}{33}%
	\makeatletter
	\providecommand \@ifxundefined [1]{%
		\@ifx{#1\undefined}
	}%
	\providecommand \@ifnum [1]{%
		\ifnum #1\expandafter \@firstoftwo
		\else \expandafter \@secondoftwo
		\fi
	}%
	\providecommand \@ifx [1]{%
		\ifx #1\expandafter \@firstoftwo
		\else \expandafter \@secondoftwo
		\fi
	}%
	\providecommand \natexlab [1]{#1}%
	\providecommand \enquote  [1]{``#1''}%
	\providecommand \bibnamefont  [1]{#1}%
	\providecommand \bibfnamefont [1]{#1}%
	\providecommand \citenamefont [1]{#1}%
	\providecommand \href@noop [0]{\@secondoftwo}%
	\providecommand \href [0]{\begingroup \@sanitize@url \@href}%
	\providecommand \@href[1]{\@@startlink{#1}\@@href}%
	\providecommand \@@href[1]{\endgroup#1\@@endlink}%
	\providecommand \@sanitize@url [0]{\catcode `\\12\catcode `\$12\catcode
		`\&12\catcode `\#12\catcode `\^12\catcode `\_12\catcode `\%12\relax}%
	\providecommand \@@startlink[1]{}%
	\providecommand \@@endlink[0]{}%
	\providecommand \url  [0]{\begingroup\@sanitize@url \@url }%
	\providecommand \@url [1]{\endgroup\@href {#1}{\urlprefix }}%
	\providecommand \urlprefix  [0]{URL }%
	\providecommand \Eprint [0]{\href }%
	\providecommand \doibase [0]{http://dx.doi.org/}%
	\providecommand \selectlanguage [0]{\@gobble}%
	\providecommand \bibinfo  [0]{\@secondoftwo}%
	\providecommand \bibfield  [0]{\@secondoftwo}%
	\providecommand \translation [1]{[#1]}%
	\providecommand \BibitemOpen [0]{}%
	\providecommand \bibitemStop [0]{}%
	\providecommand \bibitemNoStop [0]{.\EOS\space}%
	\providecommand \EOS [0]{\spacefactor3000\relax}%
	\providecommand \BibitemShut  [1]{\csname bibitem#1\endcsname}%
	\let\auto@bib@innerbib\@empty
	\bibitem [{\citenamefont {Kovrizhin}\ and\ \citenamefont
		{Chalker}(2010)}]{Kovrizhin2010}%
	\BibitemOpen
	\bibfield  {author} {\bibinfo {author} {\bibfnamefont {D.~L.}\ \bibnamefont
			{Kovrizhin}}\ and\ \bibinfo {author} {\bibfnamefont {J.~T.}\ \bibnamefont
			{Chalker}},\ }\href
	{http://journals.aps.org/prb/abstract/10.1103/PhysRevB.81.155318} {\bibfield
		{journal} {\bibinfo  {journal} {Phys. Rev. B}\ }\textbf {\bibinfo {volume}
			{81}}, \bibinfo{pages}{155318} (\bibinfo {year} {2010})}\BibitemShut {NoStop}%
	\bibitem [{\citenamefont {Kramer}\ and\ \citenamefont
		{MacKinnon}(1993)}]{Kramer1993_2}%
	\BibitemOpen
	\bibfield  {author} {\bibinfo {author} {\bibfnamefont {B.}~\bibnamefont
			{Kramer}}\ and\ \bibinfo {author} {\bibfnamefont {A.}~\bibnamefont
			{MacKinnon}},\ }\href {\doibase 10.1088/0034-4885/56/12/001} {\bibfield
		{journal} {\bibinfo  {journal} {Reports Prog. Phys.}\ }\textbf {\bibinfo
			{volume} {56}},\ \bibinfo {pages} {1469} (\bibinfo {year}
		{1993})}\BibitemShut {NoStop}%
	\bibitem [{\citenamefont {Thouless}(1972)}]{Thouless1972}%
	\BibitemOpen
	\bibfield  {author} {\bibinfo {author} {\bibfnamefont {D.~J.}\ \bibnamefont
			{Thouless}},\ }\href {\doibase 10.1088/0022-3719/5/1/010} {\bibfield
		{journal} {\bibinfo  {journal} {J. Phys. C Solid State Phys.}\ }\textbf
		{\bibinfo {volume} {5}},\ \bibinfo {pages} {77} (\bibinfo {year}
		{1972})}\BibitemShut {NoStop}%
	\bibitem [{\citenamefont {Wei{\ss}e}\ \emph {et~al.}(2006)\citenamefont
		{Wei{\ss}e}, \citenamefont {Wellein}, \citenamefont {Alvermann},\ and\
		\citenamefont {Fehske}}]{Weisse2006}%
	\BibitemOpen
	\bibfield  {author} {\bibinfo {author} {\bibfnamefont {A.}~\bibnamefont
			{Wei{\ss}e}}, \bibinfo {author} {\bibfnamefont {G.}~\bibnamefont {Wellein}},
		\bibinfo {author} {\bibfnamefont {A.}~\bibnamefont {Alvermann}}, \ and\
		\bibinfo {author} {\bibfnamefont {H.}~\bibnamefont {Fehske}},\ }\href
	{\doibase 10.1103/RevModPhys.78.275} {\bibfield  {journal} {\bibinfo
			{journal} {Rev. Mod. Phys.}\ }\textbf {\bibinfo {volume} {78}},\ \bibinfo
		{pages} {275} (\bibinfo {year} {2006})}\BibitemShut {NoStop}%
\end{thebibliography}


\onecolumngrid
\newpage
\twocolumngrid

{\bf Supplementary Material}


\emph{Numerical Methods}. First we discuss the procedure which we used to evaluate the correlators. Since all operators in the original Hamiltonian~(1) of the main text either conserve or flip charges locally, all correlators can be written as fermionic ones evolving under the single-particle Hamiltonian~(2) of the main text, and have to be averaged over all charge configurations (for our choice of a polarized initial spin-state). For example, an expectation value of the $\hat{\sigma}^z$ spin component at time $t$ after the quench reads,
\begin{multline}\label{eq: spin average}
\la 0 | \hat{\sigma}^z_{j,j+1}(t) | 0 \ra=\la 0|e^{i \hat H t}\hat{\tau}^{x}_j\hat{\tau}^x_{j+1}e^{-i \hat H t}|0\ra\\
=\frac{1}{2^{N}} \sum_{\{q_i\}=\pm1} \la \psi | e^{i H(\bar{q}_j)t}e^{-iH(\bar{q}_{j+1}) t} |\psi \ra,
\end{multline}
with the signs of the charges at locations denoted with $\bar{q}$ 
opposite to the ones appearing in equation~(2) of the main text. Other correlators of physical spins $\hat{\sigma}$ and fermions $\hat{f}$ can be represented in a similar fashion. We note that while the Hamiltonian~(2) describes a tight-binding model with an on-site disorder potential, the physical correlators in our model are in general distinct from the  correlators appearing naturally in the context of Anderson localization problems, cf.~equation~(7) of the main text.

Owing to the non-interacting form of~(2), the averages over initial states in~(7) can be expressed in terms of determinants~\cite{Kovrizhin2010}, and computed efficiently for any initial charge configuration. On the basic level we use expressions of the form
\begin{equation}\label{eq: exp det}
\la \alpha | \exp\{ i \sum_{ij} A_{ij} \hat{c}^\dag_i \hat{c}_j \} | \beta \ra,
\end{equation}
where $A$ is a Hermitian matrix, $|\alpha \ra = \hat{c}^\dag_{m_N} \cdots \hat{c}^\dag_{m_1} | vac \ra$, and $|\beta \ra = \hat{c}^\dag_{n_N} \cdots \hat{c}^\dag_{n_1} | vac \ra$; this has a determinantal expression.

To show that, see Appendix in \cite{Kovrizhin2010}, we first use the unitarity of the exponential operator to rewrite equation~\eqref{eq: exp det} as
\begin{multline}\label{eq: c tilde}
\la vac | \hat{c}_{m_1} \cdots \hat{c}_{m_N} \hat{U}_A \hat{c}^\dag_{n_N}\hat{U}_A^\dag \cdots \hat{U}_A\hat{c}^\dag_{n_1}\hat{U}_A^\dag \hat{U}_A|vac \ra \\
= \la vac | \hat{c}_{m_1} \cdots \hat{c}_{m_N} \hat{\tilde{c}}^\dag_{n_N} \cdots \hat{\tilde{c}}^\dag_{n_1} |vac \ra,
\end{multline}
where we introduced the notation $\hat{U}_A \equiv \exp\{ i \sum_{ij} A_{ij} \hat{c}^\dag_i \hat{c}_j \}$, $\hat{\tilde{c}}^\dag_j \equiv \hat{U}_A \hat{c}^\dag_{j}\hat{U}_A^\dag$, and we use $\hat{U}_A |vac\ra = |vac\ra$. Via the Baker-Hausdorff formula we further obtain
\begin{equation}
\hat{\tilde{c}}^\dag_i = \sum_j\exp\{i A^T\}_{ij} \hat{c}^\dag_j \equiv \sum_j U^T_{A,ij} \hat{c}^\dag_j,
\end{equation}
distinguishing between the operator $\hat{U}_A$, and the matrix $U_A$ by a hat. Finally, we insert this expression into~\eqref{eq: c tilde}, and use the fermionic anti-commutation relations to obtain
\begin{equation}\label{eq: det expression}
\la \alpha | \hat{U}_A |\beta \ra = \det D, \quad D_{jk} = [U_A]_{n_j m_k},
\end{equation}
with $j,k = 1,\dots N$.

This identifies possibilities for generalisations. If we have more than one unitary operator then we get
\begin{equation}
\la \alpha | \hat{U}_A \hat{U}_B \cdots |\beta \ra = \det D, \quad D_{jk} = [U_A U_B \cdots]_{n_j m_k},
\end{equation}
which simply follows from a repeated use of the Baker-Hausdorff formula. This has precisely the form needed in equation~\eqref{eq: spin average}. For the fermion correlators, we need to consider expressions such as
\begin{equation}
C_{kl} = \la \alpha | \hat{c}^\dag_k \exp\{i \sum_{ij} A_{ij}\hat{c}^\dag_i \hat{c}_j \} \hat{c}_l | \beta \ra.
\end{equation}
By commuting $\hat{c}^\dag_k$ to the left, and $\hat{c}_l$ to the right, we pick up factors $(-1)^{N-p}$ and $(-1)^{N-q}$, where $m_p = k$ and $n_q = l$, and arrive at
\begin{equation}
\begin{aligned}
C_{kl} = (-1)^{p+q}\la vac | &\hat{c}_{m_1} \cdots \hat{c}_{m_{p-1}} \hat{c}_{m_{p+1}} \cdots \hat{c}_{m_N}\\
&\times \hat{U}_A \hat{c}^\dag_{n_N} \cdots \hat{c}^\dag_{n_{q+1}} \hat{c}^\dag_{n_{q-1}} \cdots \hat{c}^\dag_{n_1} |vac \ra.
\end{aligned}
\end{equation}
This equation has the same form as~\eqref{eq: exp det}. In this case we need to remove the $q$-row and the $p$-column before taking the determinant and then multiply by the corresponding sign, i.e., we compute the $q-p$ cofactor of $D$ where $D$ has the same form as in equation~\eqref{eq: det expression}. When the matrix $D$ is invertible the final expression can be written in a simple form 
\begin{equation}
C_{kl} = D^{-1}_{lk} \det D,
\end{equation}
where $D_{jk} = [U_A]_{n_j m_k}$, $j,k = 1,\dots N$.

Our mapping allows us to reach system sizes far beyond exact-diagonalization studies. One can estimate the size of the fermionic Hilbert space at half-filling as $N^{-1/2}2^N$ with the spin degrees of freedom adding another factor of $2^N$. Instead of diagonalizing exponentially large matrices the identification of conserved charges allows us to sample uniformly from $\sim 2^N$ determinants of $N\times N$ matrices, corresponding to different charge configurations. Finally, finite-size scaling as well as exact results (up to $N=20$) show that the required number of samples for a given accuracy scales polynomially with $N$, see inset of Fig.~4(a). This allows us to access even larger system sizes than we studied in this paper. However, this is not necessary because of the finite localization length, and the results for the system sizes presented here are sufficient. Typically we sample over $10^3-10^5$ spin configurations.

To calculate the localization length for a tight-binding model with a binary disorder potential, see Fig.~3, we used a spectral formula~\cite{Kramer1993_2,Thouless1972}
\begin{equation}
\frac{1}{\lambda_{sp}} = \min \int^\infty_{-\infty} \Omega(x) \ln |E-x| \;\text{d}x,
\end{equation}
where $\Omega$ is the single-particle density of states (DOS) of the Hamiltonian in equation~(2) of the main text. The DOS was obtained using the kernel polynomial method~\cite{Weisse2006}. The basic idea is to expand the DOS as a series in Chebyshev polynomials, the recursive definition of which allows us to utilise efficient multiplication of large sparse matrices, see Ref.~\cite{Weisse2006} for more details. We used $1,000,000$ sites with $2,400$ terms in the expansion.

\begin{figure}[htb!]
	\centering
	\subfigimg[width=.38\textwidth,valign=t]{\hspace*{10pt} \textbf{(a)}}{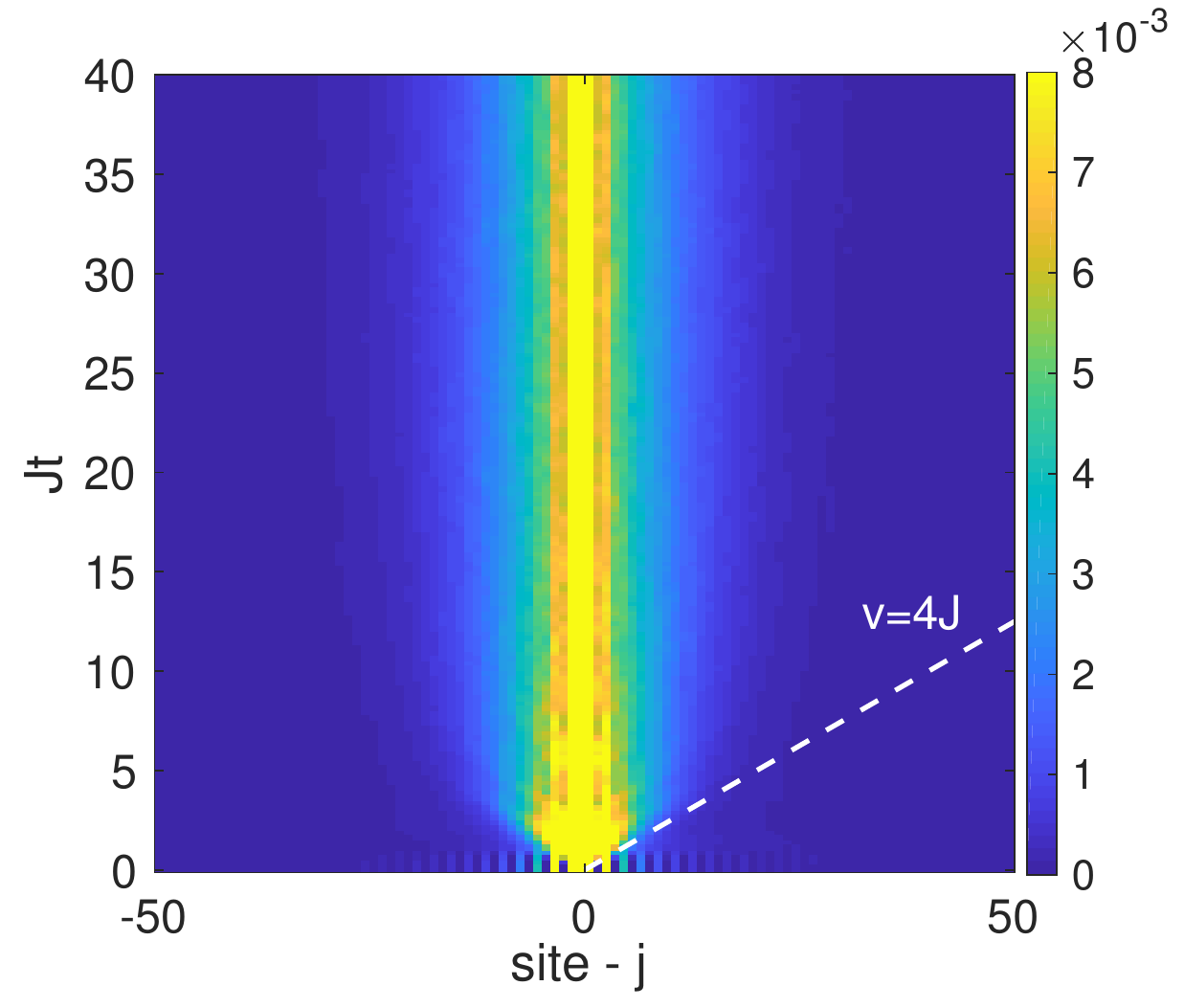}\\
	\subfigimg[width=.43\textwidth,valign=t]{\hspace*{7pt} \textbf{(b)}}{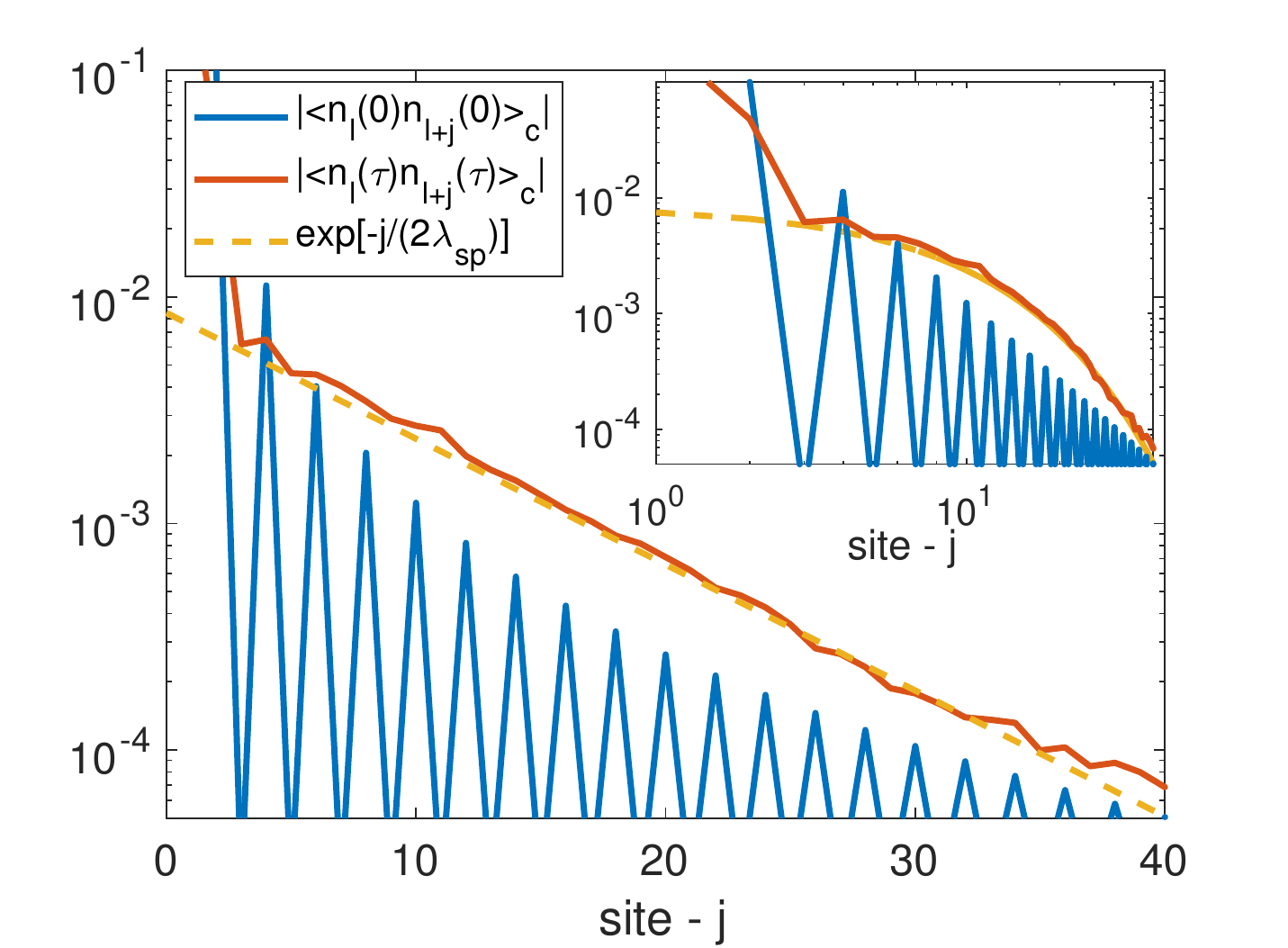}
	\caption{Connected density-density correlator from a translationally invariant Fermi-sea initial state with $h/J=0.6$, $N=200$ sites and periodic boundary conditions. (a) absolute value of the connected density-density correlator $\la 0 | \hat{n}_l(t) \hat{n}_{l+j}(t) | 0 \ra_c$ as a function of separation $j$ and time $t$. (b) semi-log plot of the spatial correlator for $t=0$ and $t=\tau=100/J$ with the exponential $\exp\{-j/(2\lambda_{sp})\}$, where $\lambda_{sp}$ is the single particle localization length. (inset) same data on a log-log plot.}\label{fig: Fermi sea}
\end{figure}

\emph{Translationally invariant initial state}. One of the initial fermion configurations that we considered was a translationally invariant Fermi-sea. We use periodic boundary conditions and initialize the spins in the z-polarized state and the fermions in the half-filled ground state of our model with $h=0$, which corresponds to a free hopping model. We then quench $h$ to a non-zero value as before.

We show the results for the connected density-density correlator in Fig.~\ref{fig: Fermi sea} where we can identify evidence of localization. In Fig.~\ref{fig: Fermi sea}(a) we see an initially linear light-cone corresponding to the Lieb-Robinson bound $v_{LR}=4J$, as for the charge density wave. This spreading is eventually suppressed and the correlators assume a stationary form. Looking at the spatial dependence for fixed times in Fig.~\ref{fig: Fermi sea}(b) we see that for this initial state has some spatial correlation that decays algebraically as can be seen in the inset. For $Jt=100$, where the distribution has approximately taken its stationary form, we observe exponential tails which matches those seen for the domain wall and the charge density wave set by the single-particle localization length.

\emph{Initial spin states: gauge transformation}. We note in the main text that the observed localization behaviour is not specific to the polarized spin state $|S\ra = |\uparrow \uparrow \uparrow \cdots \ra$. We have checked this, using exact diagonalization, for a range of initial spin states. Furthermore, it is also easy to show that any spin state that is itself a simple tensor product in the z-basis, i.e. $|S\ra = |\updownarrow \updownarrow \updownarrow \cdots \ra$, where each spin can independently be up or down, is equivalent to the polarized state through a gauge transformation.

The gauge transformation proceeds as follows. Consider a unitary operator such that $P |\updownarrow \updownarrow \updownarrow \cdots \ra = |\uparrow \uparrow \uparrow \cdots \ra$, that is a product of local operators that flips any down spin to an up spin. The operator
\begin{equation}
P = \prod_{j - \text{down spins}} \hat{\sigma}^x_j,
\end{equation}
does just that, where $j$ runs over only the spins that are down. The unitary transformation of the Hamiltonian $\widetilde{H} = P\hat{H}P^\dagger$ follows by noting $\hat{\sigma}^x \hat{\sigma}^x \hat{\sigma}^x = \hat{\sigma}^x$ and $\hat{\sigma}^x \hat{\sigma}^z \hat{\sigma}^x = - \hat{\sigma}^z$, and so equation (1) of the main text transforms to
\begin{equation}
\widetilde{H} = -J\sum_{\langle ij\rangle} s_{i,j}\hat{\sigma}^z_{i,j} \hat{f}^\dag_{i} \hat{f}_{j}
- h \sum_{i} \hat{\sigma}^x_{i-1, i} \hat{\sigma}^x_{i,i+1},
\end{equation}
where $s_{i,j} = -1$ if the spin along that bond was down and 1 if it was up. This sign, however, can be incorporated into the fermions as follows
\begin{equation}
\hat{f}_j \rightarrow \Bigg(\prod_{i\leq j} s_{i-1,i}\Bigg) \hat{f}_j,
\end{equation}
which puts $\widetilde{H}$ in exactly the same form as equation (1). For periodic boundary conditions there are two inequivalent sectors of the gauge transformation. However, in making the duality transformation we already have to explicitly restrict ourselves to one of these sectors and thus this poses no additional complication.

\end{document}